# Escalation of Commitment: A Case Study of the United States Census Bureau Efforts to Implement Differential Privacy for the 2020 Decennial Census

(Forthcoming in: In Proceedings of Privacy in Statistical Databases - PSD 2024)

Krishnamurty Muralidhar

University of Oklahoma, Norman, OK, 73019, USA (krishm@ou.edu)

Steven Ruggles

University of Minnesota and Director of IPUMS, Minneapolis, MN 55455 (ruggles@umn.edu)

## Abstract

In 2017, the United States Census Bureau announced that because of high dis-closure risk in the methodology (data swapping) used to produce tabular data for the 2010 census, a different protection mechanism based on differential privacy would be used for the 2020 census. While there have been many studies evaluating the result of this change, there has been no rigorous examination of disclosure risk claims resulting from the released 2010 tabular data. In this study we perform such an evaluation. We show that the procedures used to evaluate disclosure risk are unreliable and resulted in inflated disclosure risk. Demonstration data products released using the new procedure were also shown to have poor utility. However, since the Census Bureau had already committed to a different procedure, they had no option except to escalate their commitment. The result of such escalation is that the 2020 tabular data release offers neither privacy nor accuracy.

## Introduction

*Escalating commitment (or escalation) refers to the tendency for decision makers to persist with failing courses of action. (Brockner 1992)*

To date, there has never been a documented case of reidentification (the ability to identify a real individual based on summary data released to the public) using the decennial tabular data released to the public. Yet, in 2017, the Chief Scientist of the Census Bureau (CB) issued the following statement (Abowd, 2017): "This [database reconstruction] theorem is the death knell for public-use detailed tabulations and microdata sets as they have been traditionally prepared." No evidence to support this statement was offered at this time. This was followed by the announcement in September 2017 (Garfinkel 2017): "The 2020 Census disclosure avoidance system will use differential privacy to defend against a reconstruction attack." There was still no evidence being offered that there was even a need for a change since reconstruction does not necessarily lead to reidentification.

Starting in 2018, there has been a string of reports from CB regarding the reconstruction and reidentification experiments regarding the 2010 Census tabular data release. The culminating claim is that approximately 92% of the data is accurately reconstructed and over 75% of the respondents in the 2010 Census data have been identified (Hawes 2022). Unfortunately, CB did not release any details of the reconstruction and reidentification procedure until 2021 (Abowd 2021) which meant that the specific claims of reconstruction and reidentification could not be corroborated. In the meantime, the 2020 decennial census data collection had already been completed and the focus shifted to an assessment of the accuracy of the output from the differentially private procedure (DPP) implementation for the 2020 census data. Although a few studies have addressed the 2010 reconstruction and reidentification claims (Francis 2022; Muralidhar 2022; Muralidhar and Domingo-Ferrer 2023a; Muralidhar and Domingo-Ferrer 2023b; Ruggles 2018, 2024), a comprehensive examination of the claims is still missing.

## Reconstruction and Reidentification in 2010 Tabular Data Release

A closer examination of the claims made by the CB as it relates to reconstruction and reidentification reveals the lack of rigor in reconstruction, randomness in putative identification, and a logical fallacy in the confirmation of identification.

### Reconstruction

In 2010, Census data was collected at the household level. Apart from one question regarding type of housing, the remaining questions collect information on Person 1 (the owner/renter) followed by other residents of the household. For all residents, the information collected included Sex, Age, Race, and Ethnicity. For residents other than Person 1 (who is defined as the Householder), the relationship of the resident to Person 1 (14 categories) was also collected. The CB applied a disclosure limitation procedure (swapping) to protect vulnerable respondents, yielding a protected database. All tables were created using only the protected data. At the block level, the Census Bureau released 235 tables regarding both households and residents of the households.

A comprehensive reconstruction of both households and respondents by an adversary would involve reconstructing attribute values for every resident and using this information to assign them to specific households. The adversary can verify the accuracy of their reconstruction by simply verifying that the reconstructed data is consistent with all 235 tables released to the public. It is important to note that an adversary can reconstruct only the protected data from which the tables were created and never the original data (since this data is never used). Hence, the vulnerable individuals are still protected.

The Census Bureau attempted to reconstruct the original individual-level responses to the 2010 Census using only the published tabular data. There was no attempt to assign persons to households, and the reconstruction did not include the Relationship attribute. Thus, in each block, only (Sex, Age Group, Race, Ethnicity) are reconstructed (Hawes 2022). For a significant majority of the data, Ruggles (2018) pointed out that this reconstruction amounts to nothing more than looking up a set of tables, with absolutely no need for any optimization algorithms. Abowd et al (2023, page 19) provide an extensive description of an optimization model using integer linear programming to calculate the number of Asian Hispanic Females aged 22 on a particular block. We now illustrate how this question can be answered using the information from published tables and simple arithmetic.

For a significant majority of the blocks, such reconstruction can be performed using only tables P12A-I, P8, and P9. Tables P8 and P9 provide, at the block level, the count of all individuals and the count of non-Hispanic individuals, respectively. Tables P12A-P12G provide the count of all individuals by sex and age-group for seven race categories (White, Black/African American, American Indian or Alaska Native, Asian, Native Hawaiian or Pacific Islander, Other races, and Two or more races, respectively). Table P12H provides, by sex and age, the count of Hispanic individuals in the block. Finally, table P12I provides, by sex and age, the count of non-Hispanic White individuals. All these tables are available at https://data.census.gov/. Muralidhar (2022) showed that to the CB attempt to reconstruct individual year-of-age produced was little better than random assignment. Perhaps recognizing the difficulty of reconstructing single years of age from the published census tables, in recent publications the Census Bureau has focused on reconstructing grouped ages (e.g. 0-4, 5-9).

Table 1 provides the information from tables P8 and P9 (Part A), and data extract-ed from tables P12A-I for females in age-group 22-24 (Part B), from Block 3000, Tract 72, in New York County, New York. Table 1 (Part C) also provides the completed block counts computed using Parts A & B as follows:

(1) From P12I, the count of White, non-Hispanic females in age-group 22-24 is readily available as 43. Subtracting this count from all White females in age-group 22-24 (P12A) provides the count of White, Hispanic females.
(2) From P8 and P9, the count of Asian, Hispanic females in the block is 0. Hence, all 5 Asian females in this age-group must be non-Hispanic.
(3) From table P12H, there are 2 Hispanic females in the 22-24 age-group of which one is White. Hence, the remaining Hispanic female must belong to the "Other" race category.

Now that the table has been constructed, reconstruction is simply a matter of creating a list of these 50 respondents. All this is performed using simple arithmetic without any need for any additional computation.

It is important to note that Table 1 is not an isolated example. Every White individual can be reconstructed (over 74% of the population). In 88% of the occupied blocks (5,462,028 of 6,207,027), everyone in the block can be reconstructed. For over 90% of the population, every age group belonging to a (Block, Race) can be reconstructed. Examination of individual age-groups (as shown in Table 1) increases the proportion of the reconstructed population.

| Part A: Count of Individuals by Race (Entire Block) | | | Race | Part B: Count of Female, Age Group (22 – 24) (By Race) | | | Part C: Complete reconstruction of race and ethnicity of Female, Age group (22 – 24) | | |
|---|---|---|---|---|---|---|---|---|---|
| All (P8) | NH (P9) | H (P8 – P9) | | All | NH | H | All | NH | H |
| 474 | 447 | 27 | Total | 50 (P12) | | 2 (P12H) | 50 | **48** | 2 |
| 377 | 355 | 22 | White | 44 (P12A) | 43 (P12I) | | 44 | 43 | **1** |
| 10 | 9 | 1 | Black | 0 (P12B) | | | 0 | **0** | **0** |
| 1 | 1 | 0 | AIAN | 0 (P12C) | | | 0 | **0** | **0** |
| 80 | 80 | 0 | Asian | 5 (P12D) | | | 5 | **5** | **0** |
| 0 | 0 | 0 | NHPI | 0 (P12E) | | | 0 | **0** | **0** |
| 6 | 2 | 4 | Other | 1 (P12F) | | | 1 | **0** | **1** |
| 9 | 8 | 1 | Two or more races | 0 (P12G) | | | 0 | **0** | **0** |

**Table 1**. Reconstruction of Females, Age group (22-24), Block 3000, Tract 72, in New York County, New York (NH = Not Hispanic, H = Hispanic). Values in bold in Part C are computed from the information in Parts A and B.

CB has repeatedly claimed that reconstruction was possible only because of improvements in technology and algorithms (Abowd 2021). Our illustration above shows that this claim is demonstrably false. This study has demonstrated that for over 90% of the population, reconstruction is little more than looking up a set of tables, just as Ruggles (2018) had predicted.

It is important to note that CB researchers have recently claimed that implementing additional features "is straightforward" and provide a formulation for the relationship variable (Abowd et al 2023) This is misleading since formulating a problem does not mean that an unique solution will be found. Unlike (Sex, Age Group, Race) where there is three-way cross tabulation by block (tables P12A-I), there are no such cross tabulations for the relationship variable. Without cross tabulations, it is highly likely that the lack of constraints will result in non-unique solutions in most cases. CB have never published any results of reconstruction which includes the relationship variable.

**Putative Reidentification**

Reconstruction as defined by the CB does not necessarily pose a confidentiality risk. To demonstrate a disclosure threat, the Census Bureau matched their reconstructed data to an external commercial data source that includes people's identities.

In each block, even though the reconstruction involved four attributes (Sex, Age, Race, Ethnicity), matching with the external source was performed using only two attributes (Sex, Age). As suggested by the CB's Research & Methodology Group (RMG) (our emphasis), "such reidentification studies are performed by looking for unique combinations of variables in the microdata that are thought to be identifying, looking for externally available data sets that contain the same variables, and then linking data records in the two data sets using the linkage variables." (McKenna 2019) But the CB putative reidentification performs matching on all the reconstructed records, not just those with unique combinations of characteristics (Abowd 2021). There is a significant problem with this approach.

Consider a block which consists of 10 females in the (22-24) age-group all of whom are non-Hispanic White. The CB reidentification used only two attributes (Sex, Age) as linkage variables. As a result, these people are indistinguishable from one another, and any individual can be assigned any identity from the external source data. Because of this, all non-unique (random) matches will be dismissed as unprovable. This is precisely the reason that RMG considers only records that are uniquely identifiable based on the linkage variables. But by "looping through all the records", the new procedure used by CB contradicts RMG recommendations and treats these matches as putative identification.

At the national level, only 17.4% of the people are uniquely identifiable by (Sex, Age Group). This represents a strict upper bound on the proportion of people who can be putatively reidentified using (Sex, Age). By ignoring this criterion, CB claims putative identification at a remarkable 97%. This implies that close to 80% of the putative reidentifications are random and unprovable.

**Confirmed Reidentification**

Hawes (2022) claims that the identity of approximately 78% of the randomly iden-tified people during putative identification are "confirmed". The CB considers cases confirmed if the putative matches (where the reconstructed data match the commercial data on age, sex, and block) also match on their Protected Identification Key (PIK). The PIK is a unique identifier the CB assigns to both the original census data and the commercial data based on name, address, age, and sex. (Abowd 2021, Appendix B, para 18, page 7) According to Wagner and Layne (2014) PIK "is an anonymous identifier as unique as a SSN".

The steps in the confirmation of reidentification are described in Abowd 2021 (Appendix B, para 21, page 8). We describe these steps (including the putative reidentification steps for completeness):

(1) Putative reidentification: Match (Age, Sex) from the reconstructed data to the external commercial data.
   a. If a unique match is found, append PIK from source data to the re-constructed data.
   b. If there are no unique matches, randomly append PIK from source data to the reconstructed data.
(2) Confirmation: Compare (PIK, Age, Sex, Race, Ethnicity) in the reconstruct-ed data with original Census data. If they match, it confirms identity.

The problem with using the PIK to confirm whether the reconstructed data can be tied to real identities is that the reconstructed data do not have a PIK: the only attributes in the reconstructed data are age, sex, race, ethnicity, and block. The PIK in the putative reidentified data is appended from the commercial data.

We return to the example of the 10 (Female, 22-24, White, non-Hispanic) individuals who were randomly assigned a (Name, Address, PIK) during putative reidentification. Using the identity confirmation procedure, the identity of all 10 individuals is confirmed. But this confirmation is a fallacy. Comparison of (Age, Sex, Race, Ethnicity) for these individuals is the same and hence does not confirm identity. PIK is the only variable that uniquely confirms identity. But PIK is the identifier. This is a text-book example of circular logic fallacy.

Modifying this approach slightly takes it to its absurd but logical conclusion. Assume that the adversary only creates a list of individuals in a block but does not re-construct any of the variables using the tabular data. PIK is assigned to each person in this block in the external commercial data. Using CB procedure, the identity of every individual is confirmed, since PIK's match. All you need to confirm the identity is knowledge of their identity! As Ruggles (2024) points out, this amounts to nothing more than the fact that every person in this block has a PIK, not identity. If CB had followed RMG's recommended procedure, it is unlikely that the identity of any of the individuals would have been confirmed. Since CB did not follow the correct procedure, the confirmed reidentification results are vastly exaggerated.

## Policies and their impact on data usefulness

The policies adopted by the Census Bureau impose constraints on the implementation of the disclosure limitation method. For decennial 2010 Census data, we know that there were two specific legal requirements that were in place. First, Block level population counts for total population and voting-age population (aged 18+) must be maintained (Dajani et al 2017). Second, the data was also required to maintain consistency between person and housing tables. In 2010, these two requirements played a crucial role in the way the disclosure prevention method (data swapping) was implemented. Although CB considered alternative levels of data swapping, no relaxation of the primary requirements was ever considered.

DP-based disclosure prevention methodology for the decennial 2020 Census does not satisfy either requirement. Although initially CB promised that block level voting and non-voting age counts will be preserved (Dajani et al 2017), they later changed the policy in which these counts are preserved only at the state level (Abowd et al 2020). This is three levels of geography higher than the block level (with block-group, tract, and county being the intermediate levels). Recently, the Census Bureau also announced that the tabular data released from the 2020 Decennial Census will not maintain consistency between person and household tables. This implies that the population count for a given block, block-group, tract, and county level will be different based on person tables compared to the housing tables. It is estimated that over 10% of the data may contain such inconsistencies (Menger 2023). Some of these inconsistencies include:

> Blocks completely under water.
> Blocks with only children and no adults.
> Blocks with individuals but no households.
> Blocks with households but no individuals.

Researchers have also identified that block level data regarding minorities are quite inaccurate (Kenny et al 2021, 2024). The Census Bureau has also advised against using block level results since they are not accurate. In summary, compared to the 2010 tabular data release, the quality of the DPP output is poor.

## What about privacy?

One of the key features in the implementation of differential privacy is the selection of the privacy parameter $\varepsilon$. The smaller the value of $\varepsilon$, the greater the privacy protection and vice versa. The value $e^{\varepsilon}$ represents the privacy loss.

CB considered two alternate procedures:

(1) Bottom-up approach – where Laplace noise is added to every cell in the combination of (Block × Sex × Age × Race × Ethnicity) which results in 161,109,592,812 cells. This procedure was deemed to add too much noise to the data and hence not considered.

(2) Top-down approach - where concentrated differential privacy is employed to reduce the amount of noise added (Abowd and Hawes, 2023). This procedure does not provide strict $\varepsilon$-DP but only a relaxed version ($\varepsilon,\delta$)-DP. It also involves extensive post-processing to ensure that some basic requirements (non-negativity, integer values, etc.) are met.

Originally, CB chose the value of ε to be 4.2. Using this specification, CB generated a demonstration data set protected by the top-down procedure using 2010 data. This data was released to the public to allow users to assess the accuracy of DPP compared to the originally released data. The public reaction to the demonstration data was quite negative due to poor accuracy. As a result, CB increased the value of ε to 10.61 and released a second demonstration data set, which was also received poorly. CB increased ε to 19.41 and released a third demonstration data set. Finally, CB settled on a value of ε = 39.91 for individual demographic data.

One of the problems with focusing on the value of ε is that, by itself, it provides no meaningful assessment of the privacy risk. It represents the log of the odds ratio (ratio of two probabilities), the probability of a negative outcome (disclosure) to the probability of a positive outcome (non-disclosure). From the very beginning, the developers of differential privacy have been insistent that this ratio (the value of ε) should be small. According to Dwork (2011), "We tend to think ε of as, say, 0.01, 0.1, or in some cases, ln 2 or ln 3." We can assess the impact of this by constructing the privacy loss for different values of ε (Table 2).

| ε | 0.01 | 0.1 | 4.2 | 10.6 | 19.61 | 39.91 |
|---|------|-----|-----|------|-------|-------|
| Privacy Loss | 1.01 | 1.11 | 67 | 32860 | 325215956 | 215125935915741000 |

**Table 2**. Privacy loss for different values of ε

So how does ε of 39.91 compare to the suggested values? When ε = 0.1, the odds of negative outcome to positive outcome are 1.11:1; with ε = 39.91, the odds are 215125935915741000:1.

Referring to the use of the value of ε = 14 by Apple in one of their applications, McSherry, one of the co-inventors of differential privacy, is quoted as saying: "Anything much bigger than one is not a very reassuring guarantee. Using an epsilon value of 14 per day strikes me as relatively pointless" (Greenburg, 2017). The selection ε = 39.91 is anything but reassuring.

**Conclusions**

The decision to change the disclosure avoidance methodology from data swapping in 2010 to DPP in 2020 was predicated upon the fact that the disclosure risk from reconstruction and reidentification of the 2010 tabular data release was unacceptable. Considering that the original data was first checked for vulnerabilities before data swapping was applied to protect the vulnerable individuals, it is very surprising that this did not reveal the potential for reidentification. CB has consistently offered the explanation that "While the Census Bureau's confidentiality methodologies for the 2000 and 2010 censuses were considered sufficient at the time, advances in technology in the years since have reduced the confidentiality protection provided by data swapping." (Abowd 2021, Keller and Abowd 2023). In this paper, we have shown that this statement is provably false. Using only a few tables and simple arithmetic, we have shown that we can reconstruct (Sex, Age Group, Race, Ethnicity) for over 90% of the population. Given that the publicly released tables included a crosstabulation of age by sex by race and ethnicity for each census block, it seems improbable that CB was unaware of this simple reconstruction at the time.

This then raises the question as to why CB chose to release this data if it was vulnerable to reconstruction and subsequent reidentification. One plausible explanation is that CB concluded that the risk of reidentification, as defined by CB's own RMG, showed risk of identity disclosure to be very small. Even CB initially acknowledged that the risk of reidentification to be very small (Abowd 2018). Subsequently however, the risk of confirmed reidentification increased to as high as 78% (Hawes 2022). That massive increase in reidentification risk can be attributed exclusively to the change in the procedure used for putative and confirmed reidentification. This new procedure, which contradicts the RMG reidentification procedure, results in random reidentifications being classified as confirmed reidentifications, thereby considerably inflating the reidentification risk.

Finally, we believe that the decision by CB to adopt DPP was premature. The decision was announced in September 2017. At this time, it is unclear whether CB had thoroughly investigated the impact of implementing DPP. CB did not release the first demonstration data product until October 2019, almost two years after the announcement of the decision to adopt DPP. Given the magnitude of the change, it might have been prudent for CB to have waited for their announcement until after they had received feedback from the users. It might also have been prudent for CB to consider alternate courses of action. Absent any options, the only course of action was to escalate their commitment to DPP implementation. The result of this escalation is that the output from the 2020 decennial census is inaccurate, inconsistent, and "differential privacy delivers privacy in name only." (Dwork et al 2019)